\documentclass{PoS}
\usepackage{subfigure}
\usepackage{comment}
\def\aap{A\&A}
\def\aj{AJ}
\def\apj{ApJ}
\def\apjl{ApJL}
\def\apjs{ApJS}
\def\mnras{MNRAS}

\title{On the multiwavelength properties of several $\gamma$-ray detected narrow-line Seyfert 1 galaxies}

\ShortTitle{Multiwavelength properties of several $\gamma$-ray detected NLS1s}

\author{\speaker{Hui Yang}$^{1,2}$, Weimin Yuan$^{1,2}$, Su Yao$^{3,4}$, Hai-Wu Pan$^{1}$ and S. Komossa$^{5}$ \\
\llap{$^{1}$}Key Laboratory of Space Astronomy and Technology, National Astronomical
Observatories, Chinese Academy of Sciences, Beijing 100012, China\\
\llap{$^{2}$}School of Astronomy and Space Science, University of Chinese Academy of
Sciences, 19A Yuquan Road, Beijing 100049, China\\
\llap{$^{3}$}Kavli Institute for Astronomy and Astrophysics, Peking University,
Beijing 100871, China\\
\llap{$^{4}$}National Astronomical Observatories, Chinese Academy of Sciences, Beijing 100012, China\\
\llap{$^{5}$}Max-Planck-Institut f\"ur Radioastronomie, Auf dem H\"ugel 69, D-53121 Bonn, Germany\\
E-mail: \email{huiyang@nao.cas.cn}, \email{wmy@nao.cas.cn}}

\abstract{The $\gamma$-ray detection from several radio-loud (RL) narrow-line Seyfert 1 (NLS1) galaxies has enabled us to study powerful relativistic jets in active galactic nuclei (AGNs) with smaller black hole masses and higher accretion rates than classical blazars.
However, the sample of those $\gamma$-ray detected NLS1s available is still not large enough for a comprehensive and statistical study.
We provide a summary of our detections and follow-up studies of three $\gamma$-ray-emitting NLS1s: SDSS J211852.96$-$073227.5 with flaring $\gamma$-ray radiation $\cite{2018MNRAS.477.5127Y,2018ApJ...853L...2P}$ and SDSS J122222.55$+$041315.7 with the highest redshift by far (z$\sim1$)$\cite{2015MNRAS.454L..16Y}$, along with the prototype 1H 0323$+$342$\cite{2007ApJ...658L..13Z,2015AJ....150...23Y}$. 
And we will discuss their multiwavelength properties and variability properties, including implications from high-energy observations in $\gamma$-rays and X-rays, infrared and radio properties, and correlated variability between several wavebands.}

\FullConference{Revisiting narrow-line Seyfert 1 galaxies and their place in the Universe - NLS1 Padova\\
		9-13 April 2018 \\
		Padova Botanical Garden, Italy}

\begin{document}

\section{Introduction}

It has been ten years since the {\it Fermi} discovery of narrow-line Seyfert 1 (NLS1) galaxies as a new class of $\gamma$-ray-emitting active galactic nuclei (AGNs)$\cite{2009ApJ...699..976A}$.
Distinctly different from the paradigm that powerful relativistic jets are generally originated from typical radio-loud (RL) AGNs like blazars and radio galaxies, NLS1s are a class of AGNs with black hole (BH) masses 1$-$2 orders of magnitude smaller and Eddington ratios near or above Eddington limits inferred from their smaller widths of the broad Balmer lines.

The extreme properties of NLS1s in AGN parameter space sparked great attention from the community and systematic searches have been conducted to look for RL NLS1s with possible relativistic jets (e.g. $\cite{2006AJ....132..531K,2006ApJS..166..128Z,2008ApJ...685..801Y,2017ApJS..229...39R,2018arXiv180107234C}$, and see $\cite{2018arXiv180703666K}$ for a recent review on multiwavelength properties of RL NLS1s). 
RL NLS1s are rare (with $\sim$7\% in fraction) compared to the RL subset of normal broad-line AGNs (15$-$20\%) $\cite{2006AJ....132..531K}$. 
The further study of some very RL NLS1s (with radio loudness R > 100) has revealed their blazar-like properties, indicating the presence of relativistic jets (e.g. $\cite{2007ApJ...658L..13Z,2008ApJ...685..801Y,2003ApJ...584..147Z}$).
With the help of the {\it Fermi} $\gamma$-ray space telescope (hereinafter {\it Fermi}) and optical spectroscopic surveys like the Sloan Digital Sky Survey (SDSS), more than a dozen of $\gamma$-ray-emitting NLS1s have been detected which confirmed the existence of relativistic jets in this new class of $\gamma$-ray loud AGNs. 
In this paper, I will give a summary of three $\gamma$-ray-emitting NLS1s found and studied by our group in recent years, including SDSS J211852.96-073227.5 (hereinafter J2118$-$0732, a $\gamma$-ray flaring NLS1 galaxy $\cite{2018MNRAS.477.5127Y}$ which was also independently discovered by $\cite{2018ApJ...853L...2P}$), SDSS J122222.55$+$041315.7 (hereinafter J1222$+$0413, by far the most distant $\gamma$-ray-emitting NLS1 galaxy at redshfit of $z = 0.966$ $\cite{2015MNRAS.454L..16Y}$), and 1H 0323$+$342 (a prototype of NLS1-blazar hybrid and one of the best-studied galaxies $\cite{2007ApJ...658L..13Z,2015AJ....150...23Y}$).

\section{$\gamma$-ray variability}

So far, only a small number of $\gamma$-ray loud NLS1s have been discovered.
The comprehensive $\gamma$-ray study has been applied to several sources with the strongest $\gamma$-ray emission.

\begin{figure} 
  \centering 
  \subfigure[]{ 
    \includegraphics[width=9cm]{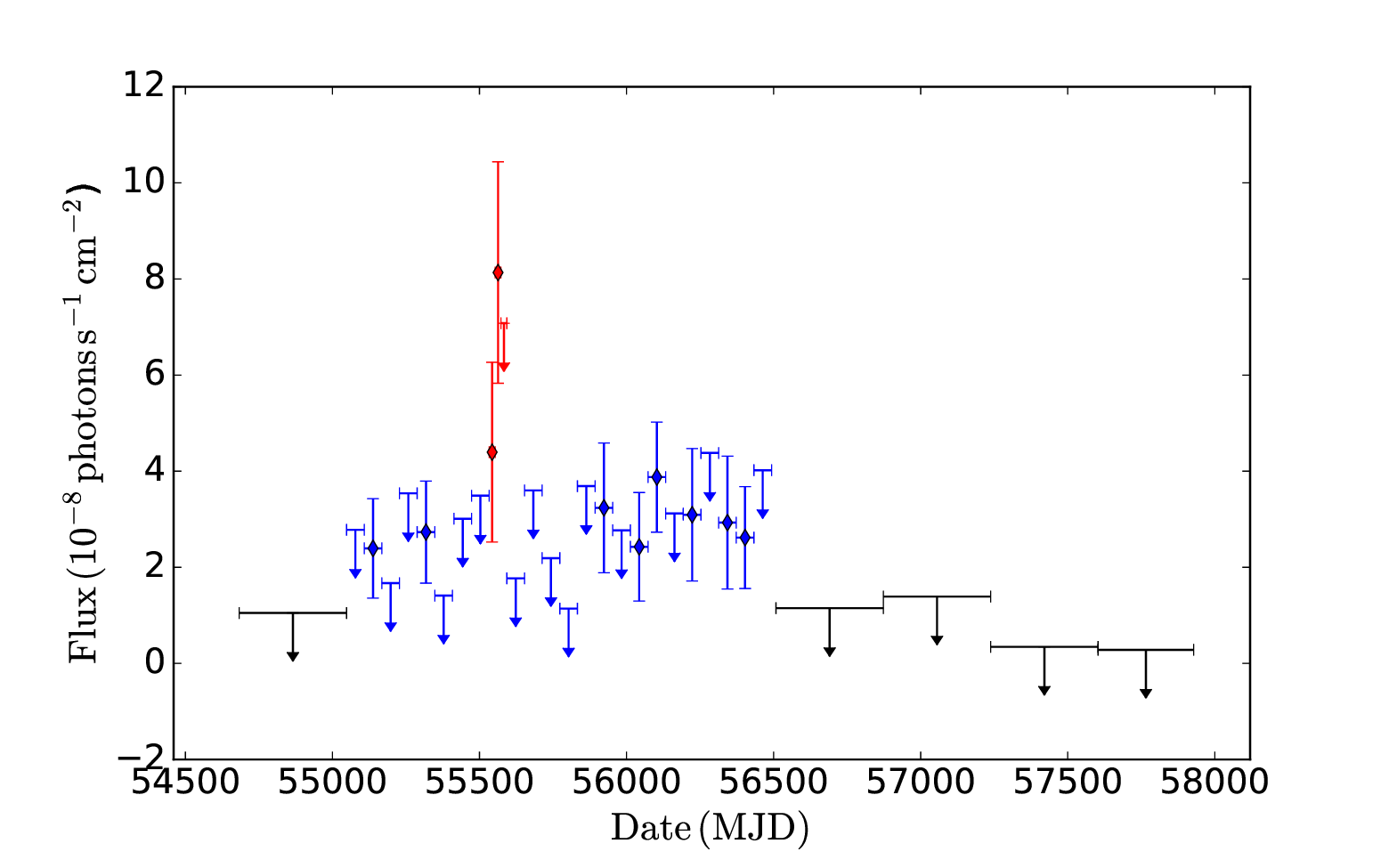} 
  } 
  \subfigure[]{ 
    \includegraphics[width=5cm]{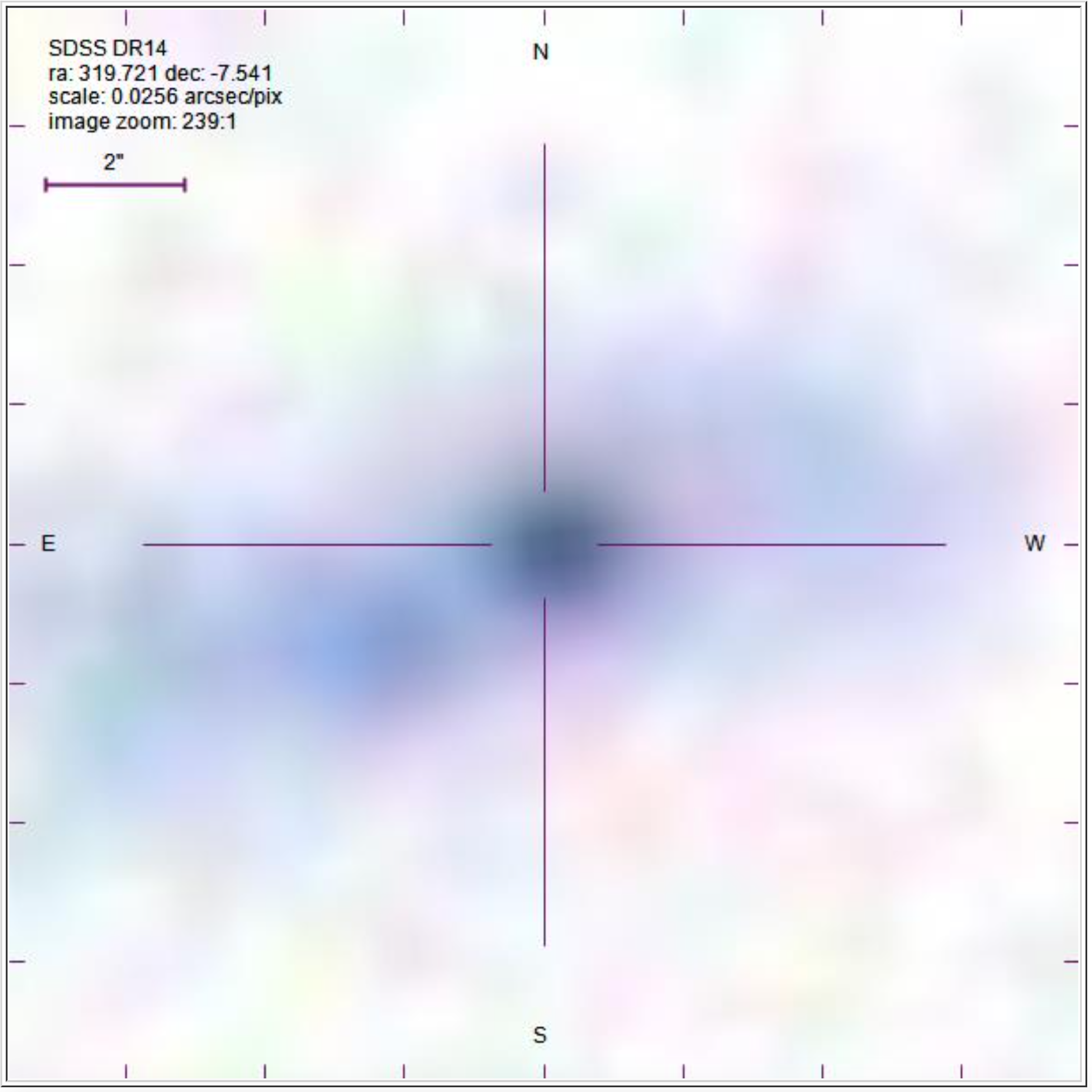} 
  } 
  \caption{$\gamma$-ray light curve (left) and SDSS image (right) of J2118$-$0732. In the left panel, $\gamma$-ray fluxes were obtained in the 0.1$-$300\,GeV band by the {\it Fermi}-LAT with 1-yr bins in black, 60-d bins in blue, and 20-d bins in red from fig. 1 in $\cite{2018MNRAS.477.5127Y}$. Diamonds refer to detections with 1$\sigma$ errors and arrows refer to 95\% upper limits. In the right panel, optical image of J2118$-$0732 were obtained from SDSS Data Release 14 database.}
  \label{fig:j2118}
\end{figure}

Our analysis from $\cite{2018MNRAS.477.5127Y}$ has shown that J2118$-$0732 was in a relatively high-flux state during 2009$-$2013 and dimmed below the detection limit of the Large Area Telescope (LAT) onboard the {\it Fermi} after 2013. 
During the active 4 years, J2118$-$0732 was detected sporadically by the {\it Fermi}-LAT, with a flaring event emerging within 20 d and quickly dropped to a lower state (see the left panel in Fig. \ref{fig:j2118}).
Rapid flares on time-scale of days have also been detected in 1H 0323$+$342 $\cite{2014ApJ...789..143P}$.
Since $\gamma$-ray flares and long-term variability are common in blazars, more $\gamma$-loud NLS1s will be detected as the {\it Fermi} continues to measure. 

\section{X-ray spectra and variability}

RL NLS1s have systematically harder X-ray spectra (with the average photon index of $\Gamma_{\rm X} = 2.0 \pm 0.5$) compared with radio-quiet NLS1s ($\Gamma_{\rm X} = 2.7$)$\cite{2015A&A...575A..13F}$, hinting perhaps at a jet origin. 
For example, the X-ray photon indices are $\Gamma_{\rm X} = 1.1 -1.5$ for J1222$+$0413 $\cite{2015MNRAS.454L..16Y}$ and $\Gamma_{\rm X} \sim 1.7 $ for J2118$-$0732 $\cite{2018MNRAS.477.5127Y}$.
1H 0323$+$342 has a softer X-ray spectrum ($\Gamma_{\rm X} \sim 1.9 $) than other $\gamma$-ray detected NLS1s, indicating that the disk/corona may make a significant contribution in the X-ray band $\cite{2015AJ....150...23Y}$. 
A soft X-ray excess has been detected in most $\gamma$-loud NLS1s that have good-quality X-ray spectra, with exceptions for J2118$-$0732 $\cite{2018MNRAS.477.5127Y}$ and PKS 2004$-$447 $\cite{2016A&A...585A..91K}$. 
Given the prevalence of soft excesses in radio-quiet NLS1s, they are likely also common in RL NLS1s. 
However, not all NLS1s have steep soft X-ray spectra, and in RL NLS1s, intrinsic soft excesses may sometimes escape detections, when the jet emission is in a high-state.

Significant X-ray variability has also been detected in $\gamma$-ray loud NLS1s at multiple time-scales. 
The X-ray flux integrated from 0.3 to 10\,keV of J2118$-$0732 dropped by a factor of $\sim$3 between the two {\it XMM-Newton} observations separated by about five months $\cite{2018MNRAS.477.5127Y}$.
A significant X-ray variability on time-scales from ks to years has been detected in 1H 0323$+$342 $\cite{2015AJ....150...23Y}$.

\section{Intraday infrared variability}

Both J2118$-$0732 and J1222$+$0413 are detected to have intraday infrared variability by studying the Wide-field Infrared Survey Explorer ({\it WISE}) data $\cite{2018MNRAS.477.5127Y,2015MNRAS.454L..16Y}$. 
Such rapid infrared variability is also detected in some other RL NLS1s $\cite{2012ApJ...759L..31J}$.
And those sources with intraday infrared variability are usually the ones with the largest radio loudness and the highest radio brightness temperature. 
The detection of such a short variability time-scale can set an upper limit on the size of the infrared-emitting region to $\leq 10^{-3}$\,pc, which is significantly smaller than the scale of the torus but consistent with that of the jet-emitting size.

\section{Radio properties}

1H 0323$+$342 is a strong radio source with a flux density of 304$-$581 mJy at 5\,GHz, and it has a flat radio spectrum up to at least 10\,GHz [$\alpha_{\rm rad} \approx$0.1 ($S_{\nu}\propto \nu^{\alpha}$)] and significant flux variations $\cite{2007ApJ...658L..13Z}$.
The polarization levels of 1H 0323$+$342 are ranging from 3\% to 5\% at multiple radio bands.
The radio spectrum of J1222$+$0413 is also very flat ($\alpha_{\rm rad} \approx$0.3) with the core flux density of 600 mJy at 1.4\,GHz and variations on time-scales from years to decade at GHz bands $\cite{2015MNRAS.454L..16Y}$.
As for J2118$-$0732, the 1.4\,GHz flux density is 96 mJy and the polarized flux density is 3.2 mJy, corresponding to a fractional polarization of $\sim 3 \%$. 
The source shows a steep spectrum at lower frequencies, however, a flat/inverted spectrum above 1.4\,GHz $\cite{2018MNRAS.477.5127Y}$.
The common feature of the flat radio spectrum at higher frequencies among RL NLS1s may also indicate a jet nature, where the flat spectrum could be the result of a superposition of several jet components $\cite{1981ApJ...243..700K}$.

These three sources have all been observed by the Very Long Baseline Array (VLBA).
The high-resolution images of 1H 0323$+$342 and J1222$+$0413 reveal a core-jet structure $\cite{2007ApJ...658L..13Z,2018arXiv180505258L}$ while that of J2118$-$0732 shows an unresolved core only\footnote{http://astrogeo.org/cgi-bin/imdb\_get\_source.csh?source=J2118-0732}, supporting the core nature.
This is consistent with the parsec-scale radio properties of a sample of RL NLS1s studied from VLBA observations $\cite{2015ApJS..221....3G}$.

\section{Correlated multiwavelength variability}

By studying the light curves at multiple wavelengths, coordinated variation patterns are found from some $\gamma$-ray NLS1s, indicating the connections among accretion discs, coronas, and jets.
With the help of intensive Swift monitoring observations, a statistically significant correlation between X-ray and ultraviolet (UV) $w2$ bands was found for 1H 0323$+$342 $\cite{2015AJ....150...23Y}$.
And a further cross-correlation method resulted in a possible time lag $\tau = 0.6_{-1.0}^{+2.7} $ day with the X-rays tentatively leading the UV.
However, the uncertainty of the time lag is quiet large and the lag is statistically consistent with zero.
For another interesting source J2118$-$0732, two {\it XMM-Newton} observations separated by about five months were proposed and carried out $\cite{2018MNRAS.477.5127Y}$.
Together with the {\it WISE} data, two broadband simultaneous spectral energy distributions (SEDs) were constructed and showed the synchronous drop from infrared to X-rays in five months.

\section{Host galaxy}

Optical observations of the host galaxies of most RL NLS1s have not yet been carried out, so we know very little about their hosts.  
Only a few $\gamma$-NLS1s have been imaged recently.
For example, 1H 0323$+$342 shows a one-armed spiral galaxy $\cite{2007ApJ...658L..13Z}$ or possibly resides in a system disturbed by merging $\cite{2008A&A...490..583A,2014ApJ...795...58L}$.
The SDSS imaging of J2118$-$0732 shows an extended galaxy structure with a disturbed morphology, possibly suggesting a recent merger (see the right panel in Fig. \ref{fig:j2118}). 
This may suggest that mergers could be important for the production of relativistic jets in RL NLS1s.

\section{Broadband SED}

The broadband SEDs of $\gamma$-ray NLS1s are representative of those blazars which exhibit double distinct peaks.
And they are usually explained by the synchrotron radiation and the inverse Compton scattering of the relativistic electrons in the jets.
Apart from the powerful jet emission, the disc/corona emission is also very strong in some RL NLS1s, contributing to the optical/UV and X-ray fluxes.
The simple one-zone leptonic jet model is often used to account for the broadband emission from GeV-NLS1 galaxies, sometimes accompanied by disc/corona emission to explain the big blue bump and softer X-ray spectrum. 

\begin{figure} 
  \centering 
  \subfigure[]{ 
    \includegraphics[width=7.6cm]{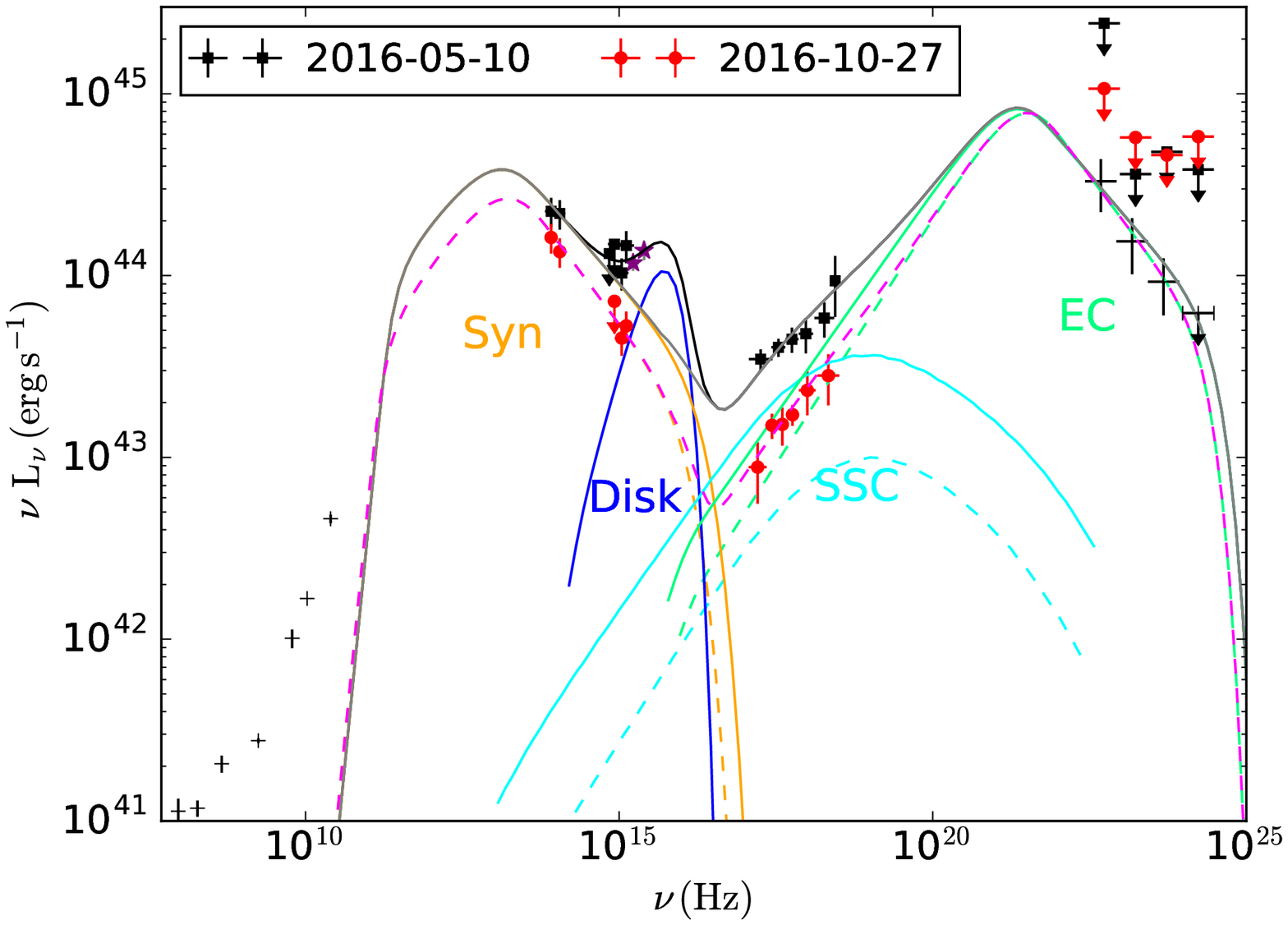} 
  } 
  \subfigure[]{ 
    \includegraphics[width=6.4cm]{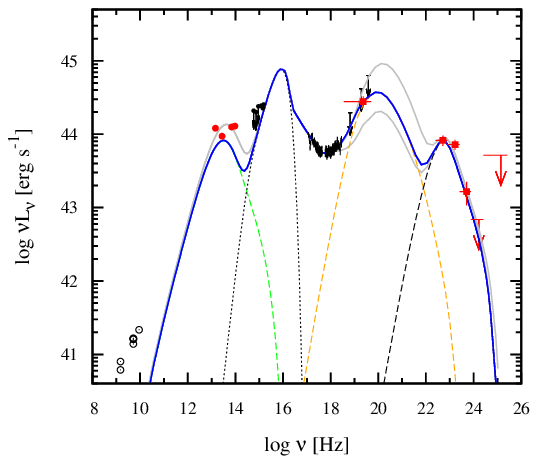} 
  } 
  \caption{Broadband SEDs of J2118$-$0732 (left) and 1H 0323$+$342 (right) and their SED modellings. In the left panel, the black squares and red dots are measurements of J2118$-$0732 at different epochs. The black and magenta lines are their best-fitting models. In the right panel, data from optical/UV to X-rays are quasi-simultaneous while others are not. The blue solid line is the best-fitting model and the gray solid lines are the contributions from the SED modelling results after the hard X-rays were varied by a factor of 2. The green, orange, and black dashed lines represent the synchrotron, synchrotron-self-Compton (SSC), and External Compton (EC) components, respectively. The black dotted line indicates the disc model.}
  \label{fig:sed}
\end{figure}

The broadband SED of 1H 0323$+$342 has been modelled in several works (e.g. $\cite{2015AJ....150...23Y,2014ApJ...789..143P,2009ApJ...707L.142A,2018MNRAS.475..404K}$), which could be well explained by the jet plus disc/corona model (see the right panel in Fig. \ref{fig:sed}). 
And the broadband SEDs of J2118$-$0732 (see the left panel in Fig. \ref{fig:sed}) and J1222$+$0413 can be well fitted with the jet+disc model $\cite{2018MNRAS.477.5127Y,2015MNRAS.454L..16Y}$.
However, the question still remains whether the dissipation region of the jet locates near the BH or farther away in the broad line region or torus.

\section{Summary}

In the last decade, we have seen a swarm of investigations of RL NLS1s, with the number of $\gamma$-ray-emitting NLS1s growing to $\sim$14 or more detected by the {\it Fermi}-LAT telescope (see table 1 of  $\cite{2018arXiv180703666K}$).
Here, we have summarized our results of three $\gamma$-ray-emitting NLS1s.
We have found blazar-like properties in these galaxies, including the long-term variability with sporadic flares in $\gamma$-rays, flat spectra at radio and X-ray bands, and rapid and correlated variability at several wavebands.
Together with the NLS1 nature from optical spectroscopy, they constitute a very interesting class of objects called ``NLS1-blazar hybrid''. 
With their extreme distributions in AGN parameter space, RL NLS1s allow us to explore the crucial questions regarding the formation and evolution of relativistic jets under smaller BH masses and higher accretion rates than classical blazars as well as the coupling mechanism of jets and accretion flows.
Further observations from the {\it Fermi}-LAT, optical and radio surveys would result in more RL NLS1s, particularly for those $\gamma$-ray loud NLS1s, and will shed more light on the jet properties by systematic studies at multiple wavelengths.

\section*{Acknowledgements}

This conference has been organized with the support of the
Department of Physics and Astronomy ``Galileo Galilei'', the 
University of Padova, the National Institute of Astrophysics 
INAF, the Padova Planetarium, and the RadioNet consortium. 
RadioNet has received funding from the European Union's
Horizon 2020 research and innovation programme under 
grant agreement No~730562. 
This conference proceeding has made use of the data from the {\it Fermi}, {\it XMM-Newton}, SDSS, {\it WISE}, and the Astrogeo VLBI FITS image database.


\providecommand{\href}[2]{#2}\begingroup\raggedright

\end{document}